\documentclass[journal]{IEEEtran}

\usepackage{amsmath,amssymb,amsfonts}
\usepackage{algorithmic}
\usepackage{algorithm}
\usepackage{graphicx}
\usepackage{textcomp}
\usepackage{xcolor}
\usepackage{cite}
\usepackage{booktabs}
\usepackage{pifont}
\usepackage{multirow}
\usepackage{array}
\usepackage{tabularx}
\usepackage{mathtools}
\usepackage{bm}
\usepackage{subcaption}
\usepackage{url}
\usepackage{balance}
\usepackage{amsthm}
\usepackage{tikz}
\usetikzlibrary{decorations.pathreplacing}

\newcommand{\R}{\mathbb{R}}
\newcommand{\Z}{\mathbb{Z}}
\newcommand{\Rq}{R_q}
\newcommand{\norm}[1]{\left\lVert #1 \right\rVert}

\allowdisplaybreaks[4]
\begin{document}

\title{Quantum-Resilient Distributed Optimization for Multi-Region Unit Commitment}

%Xinliang Dai,~\IEEEmembership{Member,~IEEE},
\author{Junhong~Liu,~\IEEEmembership{Member,~IEEE}, Qinfei~Long,~\IEEEmembership{Member,~IEEE}, Alex Pengfei Zhao,~\IEEEmembership{Member,~IEEE}, \\
	 Xianping Zhong,
	Yunfeng Li,~\IEEEmembership{Member,~IEEE}, 
    Xiaohai Dai,~\IEEEmembership{Member,~IEEE}, Francis Yunhe Hou,~\IEEEmembership{Fellow,~IEEE}
	% <-this % stops a space
	%\thanks{This work was supported in part by XXX. \textit{(Corresponding author: Francis Yunhe Hou.)}}
	
	%\textit{(Corresponding author: Francis Yunhe Hou.)}
	\thanks{Junhong Liu is with the Rausser College of Natural Resources, University of California at Berkeley, Berkeley, CA 94720, USA (e-mail: junhongliu@berkeley.edu). }
%	\thanks{Lanxin Du is with the Department of Electrical and Electronic Engineering, Imperial College London, London, UK (e-mail: dulanxin.max@gmail.com).}
	\thanks{Qinfei Long is with the Centre of Advanced Power and Autonomous Systems, Hong Kong Productivity Council, Hong Kong SAR, 999077, China (email: qflong@hkpc.org).}
	\thanks{Alex Pengfei Zhao is with the Department of Energy Science and Engineering, Stanford Doerr School of Sustainability, Stanford University, Stanford, CA 94305, USA (email: alexzhao@stanford.edu).}
%	\thanks{Xinliang Dai is with the Andlinger Center for Energy and Environment, Princeton University, Princeton, NJ 08540, United States of America (email: xinliang.dai@princeton.edu).}
	\thanks{Xianping Zhong is with the Zhejiang Institute of Modern Physics, Zhejiang University, Hangzhou, 310058, China (email: xianping.zhong@zju.edu.cn).}
	\thanks{Yunfeng Li is with the Department of Integrated Circuit, Shenzhen Polytechnic University, Shenzhen, 518055, China (e-mail: yunfengli@szpu.edu.cn).}
%	\thanks{Rong-Peng Liu is with the Department of Electrical and Computer Engineering, McGill University, Montreal, QC H3A 0E9, Canada (e-mail: rongpeng.liu@mcgill.ca).}
	\thanks{Xiaohai Dai is with the School of Computer Science and Technology, Huazhong University of Science and Technology, Wuhan, 430074, China (email: xhdai@hust.edu.cn).}
	\thanks{Francis Yunhe Hou is with the Department of Electrical and Computer Engineering, The University of Hong Kong, Hong Kong SAR, 999077, China (e-mail: {yhhou}@eee.hku.hk).} \vspace{-2.0em}}

\maketitle

\begin{abstract}
Multi-region unit commitment with reserve sharing requires coordinated optimization across jurisdictionally distinct system operators, exposing sensitive cost curves, topology, and dispatch decisions to inference attacks. The accelerating progress of quantum computing further compounds this threat. As quantum hardware matures, current classically-encrypted data flow becomes vulnerable to retrospective decryption. To enable post-quantum-secure distributed optimization, we propose a customized Benders decomposition-based approach with the global summation structure to share aggregated cuts and variables. By exploiting this structure, we further develop a multi-layer quantum-resilient secure aggregation protocol comprising additive masking for information-theoretic content privacy, affine variable transformation hiding individual sensitive data flows, and reveal-bound lattice-based zero-knowledge proofs providing resilience against active adversaries. Simulation results show that the proposed approach achieves the mean suboptimality of 0.09\%-0.22\% with lightweight computational overhead, recovers up to 51\% of system cost via inter-regional reserve sharing, and imposes no measurable cost-quality trade-off, whereas the noisy ADMM degrades monotonically under tightening privacy budgets and becomes structurally infeasible on combinatorially dense systems.
\end{abstract}


% ============================================================================
\section{Introduction}
\label{sec:intro}
% ============================================================================

\IEEEPARstart{C}{oordinated} unit commitment with reserve sharing across interconnected multi-region power systems presents a tension between optimization efficiency and data confidentiality. In deregulated electricity markets, each independent system operator (ISO) treats its generation fleet, transmission topology, and cost structures as proprietary assets whose disclosure could compromise competitive positions, expose strategic vulnerabilities, or violate regulatory mandates protecting commercially sensitive operational information. These pressures are intensifying as deeper renewable integration multiplies cross-region re-dispatch and reserve-sharing transactions~\cite{ahmadi2013multi}, increasing both the volume of inter-region data queries and the sensitivity of shared data across regions. Classical centralized approaches~\cite{kim1997coarse} require full visibility into all regional data, a mechanism that is often impractical in multi-region environments and exposes operators to single-point compromise. Decomposition methods~\cite{rahmaniani2017benders, geoffrion1972generalized, conejo2006decomposition, bragin2015convergence} preserve regional autonomy but still exchange optimization signals, such as primal variables, dual variables, or gradient information, that can reveal sensitive operational data to honest-but-curious participants observing iteratively across repeated dispatch runs~\cite{dvorkin2020differentially, han2016differentially, ruiz2013revealing}. Privacy-preserving distributed optimization that simultaneously respects regional autonomy, preserves solution quality, and resists adversarial inference is therefore a prerequisite for next-generation multi-region energy management.

The accelerating progress of quantum computing further complicates this picture. Shor's algorithm~\cite{shor1999polynomial} renders RSA, Diffie-Hellman, and elliptic-curve cryptography insecure on quantum hardware, and a harvest-now-decrypt-later adversary can record classically-encrypted inter-control-center data flows for retrospective decryption once cryptographically relevant quantum hardware matures. Existing privacy mechanisms for power-system optimization therefore face a trilemma among accuracy, computational cost, and quantum resilience. Differential privacy~\cite{dvorkin2020differentially, mak2019privacy, mak2020privacy} provides formal information-theoretic $(\varepsilon, \delta)$-guarantees, but injects calibrated noise that degrades solution quality by $\mathcal{O}(1/\varepsilon)$, which is an unacceptable trade-off for unit commitment, where suboptimality translates directly into excess fuel and start-up costs at scale. The multiplicative perturbation of consensus iterates can also render binary commitment subproblems structurally infeasible. Homomorphic encryption~\cite{lu2012eppa, wu2021privacy, lu2018privacy,si2022distributed} preserves exactness on the algebraic operations that its underlying scheme supports, but its dominant deployment model is two-party (client-server); extending it to the multi-region setting requires threshold or multi-key constructions whose ciphertext-expansion and bootstrapping overheads scale poorly with the number of parties. Meanwhile, its widely deployed Paillier and RSA forms remain Shor-vulnerable. Information-theoretic secure multi-party computation~\cite{tian2022fully, liu2023privacy} offers strong guarantees, but remains restricted to continuous optimization (e.g., OPF, economic dispatch) with communication overhead that scales quadratically in the number of parties and linearly in the circuit depth. These limitations are particularly acute for multi-region unit commitment, where the combination of discrete commitment decisions, iterative cut-passing decompositions, and limited operational solution time admits little tolerance for either accuracy degradation or super-linear protocol overhead.

The most recent attempts to address quantum resilience build on lattice problems, which has been standardized by the U.S. National Institute of Standards and Technology (NIST) for post-quantum cryptography \cite{nist2024fips203}. Ring-Learning-With-Errors (RLWE)~\cite{lyubashevsky2010ideal, regev2009lattices, peikert2016decade} provides a particularly attractive foundation, combining strong worst-case-to-average-case security reductions with efficient polynomial-ring arithmetic that admits hardware acceleration through the Number Theoretic Transform (NTT). Y.~Lu \cite{lu2025lzksa} recently demonstrated that RLWE-based zero-knowledge proofs reduce proof-generation time by three orders of magnitude compared with prior methods; however, their approach targets data aggregation rather than optimization computation, and does not address the iterative and structured data sharing that defines distributed mixed-integer optimization. To the best of our knowledge, no prior work simultaneously handles mixed-integer unit commitment, provides post-quantum security, and maintains negligible computational overhead in a decomposition approach.

This paper bridges the identified gap by exploiting a structural observation that has not, to our knowledge, been previously formalized in the privacy-preserving distributed optimization literature. The inter-regional data sharing in customized Benders decomposition takes the form of a global summation. This structure naturally motivates the secure data-sharing protocol design. We layer additive masking with an affine variable transformation that hides individual sensitive data flows from adversaries and a reveal-bound lattice-based zero-knowledge proofs for secure aggregation (RB-LZKSA), with all data transmitted over the quantum-safe channels~\cite{bos2018crystals}. The protocol preserves correctness of the underlying shared data. The contributions of this paper are as follows:
\begin{enumerate}
	\item We propose a customized Benders decomposition-based distributed optimization approach for the multi-region unit commitment with reserve sharing. This approach leverages Benders' global summation structure to share aggregated cuts and variables.
	\item We further develop a multi-layer quantum-resilient secure aggregation protocol comprising additive masking for information-theoretic content privacy, affine variable transformation that hides individual sensitive data flows, and reveal-bound lattice-based zero-knowledge proofs for secure aggregation (RB-LZKSA) providing resilience against active adversaries. The protocol operates directly on the shared data without introducing errors and brings lightweight computational overhead at most $3.8\%$ of total solve time.
	\item We validate the overall approach on the standard test systems. The proposed approach achieves post-quantum privacy without measurable cost-quality trade-off compared to its privacy-disabled counterpart, whereas the noisy ADMM degrades monotonically under tightening privacy budgets and is structurally infeasible on the IEEE~300-bus system.
\end{enumerate}
The remainder of this paper is organized as follows. Section~\ref{sec:formulation} formulates the multi-region unit commitment problem. Section~\ref{sec:benders} develops the Benders decomposition algorithm with proximal stabilization. Section~\ref{sec:privacy} presents the multi-layer post-quantum privacy approach and formalizes the structural privacy comparison with ADMM. Section~\ref{sec:results} provides comprehensive computational results, and Section~\ref{sec:conclusion} concludes with future directions.

% ============================================================================
\section{Problem Formulation}
\label{sec:formulation}
% ============================================================================

\subsection{Notation}

We use boldface lowercase $\bm{\theta}$ for vectors and uppercase $\bm{\Phi}$ for matrices. $\R^n$ denotes $n$-dimensional real space. $\Z_q$ denotes integers modulo $q$. $\Rq = \Z_q[x]/(x^n+1)$ is the cyclotomic polynomial ring. $\norm{\cdot}$ is the Euclidean norm. $\odot$ denotes the Hadamard (element-wise) product. $\xleftarrow{\$}$ denotes uniform random sampling. $\text{clip}(\cdot, a, b)$ clamps each element to $[a,b]$. $D_{\Z^n, \sigma}$ is the discrete Gaussian distribution with standard deviation $\sigma$. 

\subsection{Multi-Region Unit Commitment}

Consider a power system partitioned into $R$ regions, interconnected by $K$ tie-lines over $T$ time periods~\cite{fu2005security, padhy2004unit}. Let $\theta_{k,t} \in [\theta_{k}^{\min}, \theta_{k}^{\max}]$ denote the power flow on tie-line $k$ at period $t$, and define $\bm{\theta} \in \R^{KT}$ as the vector of all tie-line flow decisions, which are continuous. The centralized UC problem is modeled as:
\begin{subequations}
\label{eq:centralized}
\begin{align}
\min_{\bm{\theta}, \{u,v,w,p\}} \quad & \sum_{r=1}^{R} C_r(u_r, v_r, w_r, p_r) \label{eq:cent_obj} \\
\text{s.t.} \quad & (u_r, v_r, w_r, p_r) \in \mathcal{X}_r(\bm{\theta}), \quad \forall r \label{eq:cent_feas} \\
& \bm{\theta}^{\min} \leq \bm{\theta} \leq \bm{\theta}^{\max}, \label{eq:cent_zbounds}
\end{align}
\end{subequations}
where $C_r(\cdot)$ is the operating cost of region $r$ and $\mathcal{X}_r(\bm{\theta})$ is the feasible set of region $r$ parameterized by the tie-line flows.

\subsection{Regional Subproblem Formulation}

For each region $r$ with $N_r$ generators, the subproblem is formulated as below:
\begin{subequations}
\begin{align}
&C_r = \sum_{i=1}^{N_r} \sum_{t=1}^{T} \bigl[ c_{i,t}^{\text{gen}} p_{i,t} + c_i^{\text{su}} v_{i,t} + c_i^{\text{sd}} w_{i,t} + c_i^{\text{rup}} r_{i,t}^{\text{up}} + c_i^{\text{rdn}} r_{i,t}^{\text{dn}} \bigr]
\label{eq:region_cost} \\
&u_{i,t} - u_{i,t-1} = v_{i,t} - w_{i,t}, \forall i, t \geq 2 \label{eq:logic1} \\
&v_{i,t} + w_{i,t} \leq 1, \forall i, t \label{eq:logic2} \\
&P_i^{\min} u_{i,t} + r_{i,t}^{\text{dn}} \leq p_{i,t}, \forall i, t \label{eq:gen_lb} \\
&p_{i,t} + r_{i,t}^{\text{up}} \leq P_i^{\max} u_{i,t}, \forall i, t \label{eq:gen_ub} \\
&0 \leq r_{i,t}^{\text{up}} \leq R_i^{\text{up}},\quad 0 \leq r_{i,t}^{\text{dn}} \leq R_i^{\text{dn}}, \forall i, t \label{eq:res_bounds}\\
&p_{i,t} \leq P_i^{\max} u_{i,t} - (P_i^{\max} - \overline{P}_i^{\text{su}}) v_{i,t} \label{eq:su_cap} \\
&p_{i,t} \leq P_i^{\max} u_{i,t} - (P_i^{\max} - \overline{P}_i^{\text{sd}}) w_{i,t+1} \label{eq:sd_cap} \\
&p_{i,t} - p_{i,t-1} \leq R_i^{\text{up}} u_{i,t-1} + \overline{P}_i^{\text{su}} v_{i,t} - P_i^{\min} w_{i,t} \label{eq:ramp_up} \\
&p_{i,t-1} - p_{i,t} \leq R_i^{\text{dn}} u_{i,t} + \overline{P}_i^{\text{sd}} w_{i,t} - P_i^{\min} v_{i,t}  \label{eq:ramp_dn}\\
&\sum_{\tau=t}^{\min(t+U_i-1,\,T)} u_{i,\tau} \geq U_i \cdot v_{i,t}, \forall i, t \label{eq:min_up} \\
&\sum_{\tau=t}^{\min(t+D_i-1,\,T)} (1 - u_{i,\tau}) \geq D_i \cdot w_{i,t}, \forall i, t \label{eq:min_dn} \\
&\sum_{i=1}^{N_r} p_{i,t} \geq d_{r,t} - \sum_{k \in \mathcal{K}_r} \psi_{r,k} \cdot \theta_{k,t}, \forall t
\label{eq:power_bal} \\
&f_{l,t}^r = \sum_{b \in \mathcal{B}_r} \bm{\Phi}_r[l,b] \cdot \bigl( g_{b,t}^r - d_{b,t}^r + \text{imp}_{b,t}^r \bigr)
\label{eq:ptdf_flow} \\
&\bm{\Phi}_r = \bm{B}_f \cdot \bm{B}_{\text{red}}^{-1} \label{eq:ptdf_calc} \\
&-F_l^{\max} \leq f_{l,t}^r \leq F_l^{\max}, \forall l \in \mathcal{L}_r, \forall t
\label{eq:line_limits} \\
&\sum_{i=1}^{N_r} r_{i,t}^{\text{up}} \geq R_{r,t}^{\text{up}} - \gamma \sum_{k \in \mathcal{K}_r} (F_k^{\max} - |\theta_{k,t}|), \forall t \label{eq:res_up} \\
&\sum_{i=1}^{N_r} r_{i,t}^{\text{dn}} \geq R_{r,t}^{\text{dn}} - \gamma \sum_{k \in \mathcal{K}_r} (F_k^{\max} - |\theta_{k,t}|), \forall t, \label{eq:res_dn} 
\end{align}
\end{subequations}
where $u_{i,t}$, $v_{i,t}$, and $w_{i,t}$ are integer variables. \eqref{eq:region_cost} consists of five costs, i.e., energy production, startup ($v_{i,t}=1$), shutdown ($w_{i,t}=1$), and upward/downward spinning reserve procurement. \eqref{eq:logic1}-\eqref{eq:logic2} enforce the commitment logic; \eqref{eq:logic1} ensures $u_{i,t}$ rises by 1 on startup and falls by 1 on shutdown, while \eqref{eq:logic2} prevents simultaneous startup and shutdown. \eqref{eq:gen_lb}-\eqref{eq:gen_ub} bound committed unit outputs with downward and upward reserves. \eqref{eq:res_bounds} further bounds reserve provisions by the unit's ramp capability $R_i^{\rm up}$ and $R_i^{\rm dn}$. \eqref{eq:su_cap}-\eqref{eq:sd_cap} impose transition capacity limits; \eqref{eq:su_cap} caps the output in a startup period, i.e., $\bar{P}_i^{\rm su}$; \eqref{eq:sd_cap} caps the output in the period before shutdown, i.e., $\bar{P}_i^{\rm sd}$. \eqref{eq:ramp_up}-\eqref{eq:ramp_dn} jointly handle three operating transitions following \cite{carrion2006computationally}. Online-to-online enforces standard ramp limits $p_{i,t}-p_{i,t-1} \leq R_i^{\rm up}$ and $p_{i,t-1}-p_{i,t} \leq R_i^{\rm dn}$; cold start ($v_{i,t}=1$) reduces \eqref{eq:ramp_up} to $p_{i,t} \leq \bar{P}_i^{\rm su}$ and \eqref{eq:ramp_dn} to first-period ramp-down feasibility; online-to-shutdown ($w_{i,t}=1$) ensures pre-shutdown output satisfies $\bar{P}_i^{\rm sd}$, consistent with \eqref{eq:su_cap}-\eqref{eq:sd_cap}. \eqref{eq:min_up}-\eqref{eq:min_dn} enforce minimum online and offline duration; after a startup ($v_{i,t}=1$), the unit must remain committed for $U_i$ consecutive periods; after a shutdown ($w_{i,t}=1$), it must remain offline for $D_i$ periods. \eqref{eq:power_bal} requires regional generation to cover net demand, where the shared tie-line flow $\theta_{k,t}$ is adjusted by the fixed topological sign $\psi_{r,k}$, which encodes whether tie-line $k$ is an import ($\psi_{r,k}=1$, reducing local generation requirements) or export ($\psi_{r,k}=-1$, increasing them) for region $r$. \eqref{eq:ptdf_flow} and \eqref{eq:ptdf_calc} together compute internal line flows under the DC approximation; the PTDF matrix $\bm{\Phi}_r = \bm{B_f} \bm{B}_{\rm red}^{-1} \in \R^{|\mathcal{L}_r| \times |\mathcal{B}_r|}$ is constructed from branch-bus susceptance matrix $\bm{B_f}$ and the reference-bus-reduced admittance matrix $\bm{B}_{\rm red}$; line flows are then the linear combination of net bus injections weighted by $\bm{\Phi}_r[l,b]$. \eqref{eq:line_limits} enforces bidirectional thermal limits on these flows. \eqref{eq:res_up}-\eqref{eq:res_dn} require local upward and downward reserves to meet $R_{r,t}^{\rm up}=0.1\,d_{r,t}$ and $R_{r,t}^{\rm dn}=0.05\,d_{r,t}$, reduced by $\gamma\sum_k(F_k^{\max}-|\theta_{k,t}|)$, where $F_k^{\max}-|\theta_{k,t}|$ is the unused headroom on tie-line $k$ and $\gamma=0.3$ is the fraction deliverable as shared reserve.
\vspace{-0.4em}

%\begin{remark}[Model Assumptions]
%The formulation employs DC power flow via PTDF (no transmission losses), consistent with standard UC practice~\cite{carrion2006convex, stott2009dc}. Spinning reserve requirements at 10\% (upward) and 5\% (downward) of demand implicitly enforce N-1 generator contingency security per NERC BAL-002 standards~\cite{wood2013}. The reserve sharing factor $\gamma = 0.3$ represents a conservative estimate of the fraction of unused tie-line capacity deliverable within the 10-minute reserve response window. These assumptions are consistent with the UC literature and do not affect the privacy framework, which is agnostic to the subproblem formulation.
%\end{remark}
% ============================================================================
\section{Customized Benders Decomposition Algorithm}
\label{sec:benders}
% ============================================================================

\subsection{Decomposition Structure}

Benders decomposition method~\cite{geoffrion1972generalized, papavasiliou2014applying} is customized, and it decomposes the optimization problem~\eqref{eq:centralized} into a master problem over tie-line flows and $R$ regional subproblems.

{Master problem:}
\begin{subequations}
\label{eq:master}
\begin{align}
\min_{\bm{\theta}, \eta} \quad & \eta + \frac{\mu}{2} \norm{\bm{\theta} - \hat{\bm{\theta}}}^2 \label{eq:master_obj} \\
\text{s.t.} \quad & \eta \geq \alpha^{(j)} + (\bm{\beta}^{(j)})^\top \bm{\theta}, \quad \forall j \in \mathcal{J} \label{eq:master_cuts} \\
& \bm{\theta}^{\min} \leq \bm{\theta} \leq \bm{\theta}^{\max}, \label{eq:master_bounds}
\end{align}
\end{subequations}
where $\eta$ is the epigraph variable, $\mathcal{J}$ is the set of aggregated optimality cuts, $\hat{\bm{\theta}}$ is the stability center (incumbent feasible solution), and $\mu>0$ is the proximal weight. The quadratic term serves as the regularization in the proximal bundle method~\cite{kiwiel1990proximity}, which prevents the oscillation between extreme points that causes cutting planes to converge slowly in high dimensions~\cite{rahmaniani2017benders}. This quadratic term is only included in Phase 2 in Algorithm 1.

{Regional Subproblem} (for given $\bar{\bm{\theta}}$):
\begin{equation}
C_r(\bar{\bm{\theta}}) = \min_{(u_r,v_r,w_r,p_r) \in \mathcal{X}_r(\bar{\bm{\theta}})} C_r(u_r, v_r, w_r, p_r).
\label{eq:subproblem}
\end{equation}
%\begin{definition}[Serious and null steps~\cite{kiwiel1990proximity}]
	\label{def:serious_null}
	
{\bf{Definition 1}} (Adaptive Steps~\cite{kiwiel1990proximity}). Let $\bm{\theta}^{\text{trial}}$ solve~\eqref{eq:master_obj}-\eqref{eq:master_bounds} and let $\rho = \bigl[f(\hat{\bm{\theta}}) - f(\bm{\theta}^{\text{trial}})\bigr]/\bigl[f(\hat{\bm{\theta}}) - \eta^{*}\bigr]$ be the descent ratio, where $f(\bm{\theta}) = \sum_r C_r(\bm{\theta})$ and $\eta^{*}$ is the model value at the trial, and the threshold is fixed as $\rho_{s} = 0.05$.
	\begin{itemize}
		\item[(i)] \emph{Serious step} ($\rho \geq \rho_{s}$): the model predicted the descent reliably; set $\hat{\bm{\theta}} \!\leftarrow\! \bm{\theta}^{\text{trial}}$, $\mu \leftarrow \max(\mu_{\min},\,\mu/2)$.
		\item[(ii)] \emph{Null step} ($\rho<\rho_{s}$): the model was over-optimistic; keep $\hat{\bm{\theta}}$ unchanged, add the new cut, and tighten $\mu \leftarrow \min(\mu_{\max},\,1.5\mu)$.
	\end{itemize}
\subsection{Aggregated Cut Generation}
\label{sec:cuts}
At each iteration, region $r$ solves its LP relaxation at $\bar{\bm{\theta}}$ and extracts the power-balance dual $\pi_{r,t}$ from~\eqref{eq:power_bal} and the line-flow duals $\mu_{r,l,t}^{\pm}$ from~\eqref{eq:line_limits}. The per-region cut slope is as below:
\begin{equation}
\beta_{r}[k,t] = -\psi_{r,k}\Bigl[\pi_{r,t} + \sum_{l \in \mathcal{L}_r} \bm{\Phi}_r[l, b_k^r]\,(\mu_{r,l,t}^+ + \mu_{r,l,t}^-)\Bigr],
\label{eq:beta_r}
\end{equation}
where $\bm{\Phi}_r[l, b_k^r]$ quantifies the marginal change in active power flow on internal line $l \in \mathcal{L}_r$ resulting from a unit injection at the bus $b_k^r$ at which tie-line $k$ connects to region $r$. $\psi_{r,k}\in\{-1, 1\}$ is the topological sign from~\eqref{eq:power_bal} and $b_k^r$ is the bus at which tie-line $k$ connects to region $r$. The two terms quantify, respectively, the marginal value of an additional import on regional generation cost and its weighted impact on internal line-flow constraints. The cut intercept is given as:
\begin{equation}
\alpha_r = \text{LP}_r(\bar{\bm{\theta}}) - \bm{\beta}_r^\top \bar{\bm{\theta}}.
\label{eq:alpha_r}
\end{equation}
The aggregated cut is assembled by the transform-then-aggregate approach. Each region first maps $(\alpha_r,\bm{\beta}_r)$ through its local affine transform as in equation~\eqref{eq:cut_transform} to obtain the $\bm{y}$-space pair $(\tilde{\bm{a}}_r,\tilde{b}_r)$, and the coordinator securely sums
\begin{equation}
\tilde{\bm{a}}^{(j)} = \sum_{r=1}^{R} \tilde{\bm{a}}_r, \quad \tilde{b}^{(j)} = \sum_{r=1}^{R} \tilde{b}_r.
\label{eq:agg_cut}
\end{equation}
Linearity of the affine map ensures that $(\tilde{\bm{a}}^{(j)},\tilde{b}^{(j)})$ coincides with the transform of the $\bm{\theta}$-space sum $(\sum_r\alpha_r,\sum_r\bm{\beta}_r)$, so the cut is mathematically identical to its non-private counterpart. Detailed algorithm design is presented in Algorithm \ref{alg:benders}.

\begin{algorithm}[t]
\caption{PQ-Secure Customized Benders Decomposition}
\label{alg:benders}
\begin{algorithmic}[1]
\STATE \textbf{Initialize:} Pairwise Kyber-1024 key agreement $\rightarrow \{\kappa_{r,r'}\}$
\STATE Each region $r$ derives $(\bm{M}_r, \bm{r}_r)$ from pairwise keys in ~\eqref{eq:M_rows}; compute $\bm{M}_\eta$, $\tilde{\bm{c}}$, $\bm{\delta}$ via SecureAgg
\STATE Build transformed master QP in $\bm{y}$-space; precompute $\bm{Q} = \bm{M}_{1:KT}^\top \bm{M}_{1:KT}$
\STATE Seed cut pool via $n_p$ probes; set $\hat{\bm{\theta}}$, $\text{LB} \leftarrow -\infty$; $\Lambda \leftarrow 20$; $w\leftarrow \{0.2, 0.4, 0.6, 0.8\}$
\STATE \textbf{Phase~1: Cutting Plane:}
\FOR{$j = 1, \ldots$ until gap $\leq \epsilon_1$ or gap $\leq \epsilon$}
    \STATE Solve $\bm{y}$-space master and obtain $\bm{y}^*$ without regularization term; send $\bm{y}^*$ to each region via PQ channel
    \STATE Each region $r$: recover $\bm{\theta}^*_{\mathcal{K}_r} = \bm{M}_r \bm{y}^* + \bm{r}_r$
    \STATE \textbf{Parallel:} each region $r$ solves MIP/LP at $\bm{\theta}^*_{\mathcal{K}_r}$ and obtain $(C_r, \alpha_r, \bm{\beta}_r)$
    \STATE Each region $r$: transform cut and obtain $(\tilde{\bm{a}}_r, \tilde{b}_r)$ via~\eqref{eq:cut_transform}
    \STATE $(\tilde{\bm{a}}, \tilde{b}, C) \leftarrow \textsc{SecureAgg}\bigl(\{(\tilde{\bm{a}}_r, \tilde{b}_r, C_r)\}\bigr)$ \COMMENT{Alg.~\ref{alg:secure_agg}}
    \STATE Add $\tilde{\bm{a}} \bm{y} \leq \tilde{b}$ to master; update UB$=C$
\ENDFOR
\STATE \textbf{Phase~2: Proximal Bundle:}
\STATE Set stability center $\hat{\bm{\theta}}$, proximal weight $\mu \leftarrow \mu_0$
\FOR{$j = 1, \ldots$ until gap $\leq \epsilon$ or time/stall limit}
    \STATE Solve $\min_{\bm{y}} \tilde{\bm{c}}^\top \bm{y} + (\mu/2)\bm{y}^\top\bm{Q}\bm{y} + \bm{g}^\top\bm{y}$; distribute $\bm{y}^*$
    \STATE Generate multiple cuts at $\bm{\theta}^*_{\mathcal{K}_r}$ via $w\bm{\theta}^*_{\mathcal{K}_r} + (1-w)\bm{\theta}^{\text{lb}}_{\mathcal{K}_r}$
    \STATE Transform-then-SecureAgg each cut
    \STATE Serious/null step decision; adapt $\mu$
\ENDFOR
\STATE \textbf{Phase~3: Cut Saturation:}
\FOR{$j = 1, \ldots$ until gap $\leq \epsilon$ or LB stalls $\Lambda$ iters}
    \STATE Generate multiple cuts emphasizing $\bm{\theta}^{\text{lb}}$-region
    \STATE Transform-then-SecureAgg; update bounds
\ENDFOR
\RETURN $\bm{\theta}^{\text{best}}_{\mathcal{K}_r}$ per region, $[\text{LB}, \text{UB}]$
\end{algorithmic}
\end{algorithm}

% ============================================================================
\section{Post-Quantum Secure Aggregation}
\label{sec:privacy}
% ============================================================================

\subsection{Threat Model}
\label{sec:threat}

We adopt a threat model that combines active malicious behavior for parties under the adversary's control with honest behavior for parties outside its control. This is a generalization of the honest-but-curious (semi-honest) model in privacy-preserving optimization~\cite{tian2022fully, wu2021privacy}, where privacy holds against arbitrary observation, while integrity is established against active deviation. The adversary is computationally bounded and may be quantum-capable (i.e., able to run Shor's algorithm~\cite{shor1999polynomial}). The adversary may control the coordinator and any subset of regions $\mathcal{C}$ with $|\mathcal{C}|\leq R-2$. Parties under its control may behave maliciously, i.e., modify or inject any false date at any layer of the protocols.

\begin{figure*}[t]
	\centering
	\includegraphics[width=0.7\textwidth]{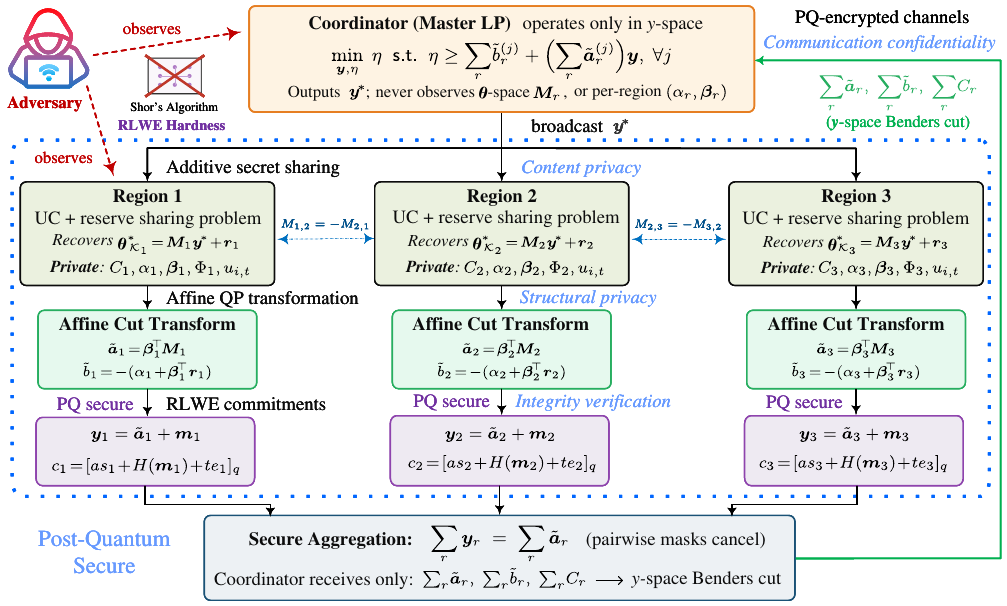}
	\caption{Post-quantum secure aggregation protocol with region-specific channels. Each region's private data pass through four cryptographic layers. Cuts are transformed per-region using the partial affine map $\bm{M}_r$ and then securely aggregated in $\bm{y}$-space. Each region receives $\bm{y}^*$ via its dedicated PQ-encrypted channel and recovers only its own tie-line flows $\bm{\theta}^*_{\mathcal{K}_r}$.}
	\label{fig:privacy_arch}
\vspace{-1em}
\end{figure*}

\subsection{Additive Secret Sharing}
\label{sec:additive_sharing}

Continuous quantities (costs and gradient components) are aggregated at arbitrary precision via pairwise antisymmetric masking. Each pair of regions $(r, r')$ shares a Gaussian mask:
\begin{equation}
M_{r,r'} \sim \mathcal{N}(0, \sigma_m^2), \quad M_{r',r} = -M_{r,r'},
\label{eq:masks}
\end{equation}
from which region $r$ forms its aggregate mask $m_r = \sum_{r' \neq r} M_{r,r'}$. The antisymmetry yields the global cancellation:
\begin{equation}
\sum_{r=1}^{R} m_r = \sum_{r < r'} (M_{r,r'} + M_{r',r}) = 0.
\label{eq:mask_cancel}
\end{equation}
Each region sends $y_r = x_r + m_r$, and the coordinator recovers
\begin{equation}
\sum_{r=1}^{R} y_r = \sum_{r=1}^{R}(x_r + m_r) = \sum_{r=1}^{R} x_r.
\label{eq:agg_correct}
\end{equation}
With $\sigma_m \gg |x_r|$, the masked value $y_r$ is statistically independent of $x_r$. Vector quantities (e.g.,\ cut slopes $\bm{\beta}_r \in \R^{KT}$) are masked elementwise with $M_{r,r'} \sim \mathcal{N}(\bm{0}, \sigma_v^2 \bm{I})$.

\subsection{RB-LZKSA: Reveal-Bound Commitment Verification}
\label{sec:rlwe}

Additive masking provides information-theoretic content privacy, but a malicious coordinator could substitute $\widehat{y}_r$ values during aggregation and bias the result. We deal with this issue by designing the RB-LZKSA, which is a reveal-bound extension of the LZKSA~\cite{lu2025lzksa} that augments the BDLOP lattice-based commitment scheme~\cite{baum2018more} with a Lyubashevsky-style non-interactive zero-knowledge proof~\cite{lyubashevsky2010ideal}. Privacy is preserved by the rejection-sampling-based zero-knowledge property of the underlying $\Sigma$-protocol~\cite{nist2024fips203}.

A public polynomial $\bm{a}\xleftarrow{\$}\Rq$ is shared. For every unordered region pair $(r,j)$ with $r<j$, the two regions run a Kyber-1024 KEM~\cite{bos2018crystals} to obtain a 256-bit shared seed $\sigma_{r,j} = \mathsf{KA.Agree}(sk_r, pk_j) = \mathsf{KA.Agree}(sk_j, pk_r)$, which is the key-agreement scheme. From this seed both regions derive, with domain-separated SHAKE-256 labels, a mask polynomial $\bm{m}^\circ_{r,j}\in\Rq$ and a commitment-randomness polynomial $\bm{s}^\circ_{r,j}\in\Rq$. Under the canonical order $r<j$, we can design:
\begin{equation}
\begin{aligned}
\text{region } r:\quad & \bm{m}_{r,j} = +\bm{m}^\circ_{r,j},\ \ \bm{s}_{r,j}=+\bm{s}^\circ_{r,j}, \\
\text{region } j:\quad & \bm{m}_{j,r} = -\bm{m}^\circ_{r,j},\ \ \bm{s}_{j,r}=-\bm{s}^\circ_{r,j},
\end{aligned}
\label{eq:antisym_split}
\end{equation}
so $\bm{m}_{r,j}+\bm{m}_{j,r}=\bm{0}$ and $\bm{s}_{r,j}+\bm{s}_{j,r}=\bm{0}$ hold by construction. Each region's aggregate quantities are $\bm{m}^*_r=\sum_{j\neq r}\bm{m}_{r,j}$ and $\bm{s}^*_r=\sum_{j\neq r}\bm{s}_{r,j}$, with global cancellation $\sum_r\bm{m}^*_r=\bm{0}$ following by telescoping. The scalar mask of Section~\ref{sec:additive_sharing} is realized at this layer as $M_{r,r'}=\mathrm{coef}_0(\bm{m}_{r,r'})$, where $\mathrm{coef}_0:\Rq\to\Z_q$ extracts the constant coefficient; the per-region scalar mask used in the reveal $y_r=x_r+m_r$ is $m_r=\mathrm{coef}_0(\bm{m}^*_r)$, which by linearity satisfies $\sum_r m_r=0$ in $\Z_q$. The remaining $n-1$ polynomial coefficients carry the algebraic structure required by the BDLOP commitment scheme. Each region publishes a BDLOP-style commitment for every neighbor:
\begin{equation}
\bm{c}_{r,j}=\bigl[\,\bm{a}\bm{s}_{r,j}+[\bm{m}_{r,j}]_t+\bm{e}^{(1)}_{r,j}\,\bigr]_q,
\label{eq:lzksa_commit}
\end{equation}
with $\bm{e}^{(1)}_{r,j}\sim D_{\Z^n,\sigma}$, $\|\bm{e}^{(1)}_{r,j}\|_\infty\leq\beta$, and $\bm{e}^*_r=\sum_{j\neq r}\bm{e}^{(1)}_{r,j}$. By linearity of the encoding, the verifier obtains the per-region homomorphic aggregate $C_r=\sum_{j\neq r}\bm{c}_{r,j}$, which opens to $(\bm{s}^*_r,\bm{m}^*_r,\bm{e}^*_r)$ with $\|\bm{e}^*_r\|_\infty\leq(R-1)\beta$. Alongside the masked reveal $y_r=x_r+m_r$, region $r$ produces a non-interactive proof $\pi_r$ that $y_r$ is consistent with $C_r$. The prover samples blinding $\bm{u}\sim D_{\Z^n,\sigma_u}$, $\bm{w}\xleftarrow{\$}\Rq$, $\bm{v}\sim D_{\Z^n,\sigma_v}$, $\rho\sim D_{\Z,\sigma_\rho}$ and computes the auxiliaries as below:
\begin{equation}
A_r=[\bm{a}\bm{u}+[\bm{w}]_t+\bm{v}]_q,\quad b_r=\rho+\mathrm{coef}_0(\bm{w}).
\label{eq:rb_aux}
\end{equation}
The Fiat-Shamir scalar challenge is as below:
\begin{equation}
\chi_r=H\!\bigl(\bm{c}_{r,1},\ldots,\bm{c}_{r,R-1},\,y_r,\,A_r,\,b_r,\,\text{round-id}\bigr)\in\mathcal{C}\equiv\Z_q\setminus\{0\},
\label{eq:rb_challenge}
\end{equation}
where $H$ is instantiated by SHAKE-256~\cite{nist2024fips203}. The public transcript is hashed and interpreted as a big-endian integer, reduced modulo $q-1$, and incremented by $1$ to land in $\mathcal{C}=\{1,\ldots,q-1\}$. The responses are then as below:
\begin{equation}
\begin{aligned}
\tilde{\bm{s}}_r&=\chi_r\bm{s}^*_r+\bm{u}, &\tilde{\bm{m}}_r&=\chi_r\bm{m}^*_r+\bm{w},\\
\tilde{\bm{e}}_r&=\chi_r\bm{e}^*_r+\bm{v}, &\tilde{x}_r&=\chi_r x_r+\rho,
\end{aligned}
\label{eq:rb_resp}
\end{equation}
with rejection sampling on $(\tilde{\bm{s}}_r,\tilde{\bm{e}}_r,\tilde{x}_r)$ to ensure the response distribution is independent of the witness. Soundness $1/|\mathcal{C}|$ per shot is amplified to negligibility by $\lambda$ parallel repetitions. The published proof is $\pi_r=(A_r,b_r,\tilde{\bm{s}}_r,\tilde{\bm{m}}_r,\tilde{\bm{e}}_r,\tilde{x}_r)$. The verifier homomorphically computes $C_r$, re-derives $\chi_r$, and accepts if the following verification are passed:
\begin{align}
\bigl[\bm{a}\tilde{\bm{s}}_r+[\tilde{\bm{m}}_r]_t+\tilde{\bm{e}}_r\bigr]_q&=\chi_r\, C_r+A_r,\label{eq:rb_v1}\\
\chi_r\, y_r+b_r&=\tilde{x}_r+\mathrm{coef}_0(\tilde{\bm{m}}_r),\label{eq:rb_v2}
\end{align}
together with the standard response-norm bounds $\|\tilde{\bm{s}}_r\|,\|\tilde{\bm{e}}_r\|,|\tilde{x}_r|\leq G$ from the rejection-sampling parameters. Equation~\eqref{eq:rb_v1} verifies that $C_r$ admits a short opening consistent with the response (commit-time integrity), and equation~\eqref{eq:rb_v2} ties $y_r$ to the scalar component of the committed mask aggregate (reveal-time integrity). Substituting~\eqref{eq:rb_resp} into the verifier's left-hand sides, equation~\eqref{eq:rb_v1} reduces to
$\chi_r(\bm{a}\bm{s}^*_r+[\bm{m}^*_r]_t+\bm{e}^*_r)+(\bm{a}\bm{u}+[\bm{w}]_t+\bm{v}) = \chi_r C_r + A_r$
by $\Z_q$-linearity of $[\cdot]_t$, and equation~\eqref{eq:rb_v2} reduces to
$(\chi_r x_r+\rho)+(\chi_r m_r+\mathrm{coef}_0(\bm{w}))=\chi_r y_r+b_r$
since $\mathrm{coef}_0$ commutes with scalar multiplication. Algorithm~\ref{alg:secure_agg} presents the RB-LZKSA verification for the secure-aggregation flow.

\begin{algorithm}[t]
\caption{Post-Quantum Aggregation with RB-LZKSA}
\label{alg:secure_agg}
\begin{algorithmic}[1]
\REQUIRE Per-region values $\{x_r\}_{r=1}^R$ (e.g.\ $\bm{y}$-space cut pieces $(\tilde{\bm{a}}_r,\tilde{b}_r)$ from~\eqref{eq:cut_transform}, regional costs $C_r$, or gradients $\nabla C_r$), public polynomial $\bm{a}\in\Rq$
\ENSURE Aggregated sum $X = \sum_{r=1}^R x_r$, or $\bot$ if any check fails
\STATE \textbf{Key Agreement (one-time per session):}
\FOR{each pair $(r,j)$ with $r < j$}
    \STATE $\sigma_{r,j} \leftarrow \mathsf{KA.Agree}(sk_r, pk_j)$
\ENDFOR
\STATE \textbf{Antisymmetric PRG Split (offline):}
\FOR{each pair $(r,j)$ with $r < j$}
    \STATE $(\bm{m}^\circ_{r,j},\,\bm{s}^\circ_{r,j}) \leftarrow \bigl(\text{SHAKE256}(\sigma_{r,j}\,\|\,\texttt{m},\texttt{s})\bigr)$
    \STATE Region $r$: $\bm{m}_{r,j}{\leftarrow}+\bm{m}^\circ_{r,j}$,\ \ $\bm{s}_{r,j}{\leftarrow}+\bm{s}^\circ_{r,j}$
    \STATE Region $j$: $\bm{m}_{j,r}{\leftarrow}-\bm{m}^\circ_{r,j}$,\ \ $\bm{s}_{j,r}{\leftarrow}-\bm{s}^\circ_{r,j}$
\ENDFOR
\STATE Each region computes $\bm{m}^*_r=\sum_{j\neq r}\bm{m}_{r,j}$, $\bm{s}^*_r=\sum_{j\neq r}\bm{s}_{r,j}$, $m_r=\mathrm{coef}_0(\bm{m}^*_r)$
\STATE \textbf{Commit phase (online):}
\FOR{each region $r$ and each neighbor $j\neq r$}
    \STATE Sample $\bm{e}^{(1)}_{r,j} \sim D_{\Z^n,\sigma}$
    \STATE $\bm{c}_{r,j} \leftarrow [\bm{a}\bm{s}_{r,j} + [\bm{m}_{r,j}]_t + \bm{e}^{(1)}_{r,j}]_q$ \COMMENT{Eq.~\eqref{eq:lzksa_commit}}
    \STATE Broadcast $\bm{c}_{r,j}$
\ENDFOR
\STATE Each region computes $\bm{e}^*_r=\sum_{j\neq r}\bm{e}^{(1)}_{r,j}$
\STATE \textbf{Reveal phase with RB-LZKSA (online):}
\FOR{each region $r$}
    \STATE $y_r \leftarrow x_r + m_r$
    \STATE Sample $\bm{u},\bm{w},\bm{v},\rho$ from the prescribed blinding distributions
    \STATE Compute $A_r,b_r$ via~\eqref{eq:rb_aux}
    \STATE Compute $\chi_r = H(\bm{c}_{r,1},\ldots,\bm{c}_{r,R-1},y_r,A_r,b_r,\text{round-id})$
    \STATE Compute responses $(\tilde{\bm{s}}_r,\tilde{\bm{m}}_r,\tilde{\bm{e}}_r,\tilde{x}_r)$ via~\eqref{eq:rb_resp}; rejection sample
    \STATE Publish $(y_r,\,\pi_r=(A_r,b_r,\tilde{\bm{s}}_r,\tilde{\bm{m}}_r,\tilde{\bm{e}}_r,\tilde{x}_r))$
\ENDFOR
\STATE \textbf{RB-LZKSA Verification:}
\FOR{each region $r$}
    \STATE $C_r \leftarrow \sum_{j\neq r}\bm{c}_{r,j}$; recompute $\chi_r$
    \IF{\eqref{eq:rb_v1} fails OR \eqref{eq:rb_v2} fails OR any response-norm bound is exceeded}
        \STATE \textbf{return} $\bot$ \COMMENT{region $r$'s contribution rejected}
    \ENDIF
\ENDFOR
\STATE Coordinator computes $X = \sum_{r=1}^R y_r = \sum_{r=1}^R x_r$
\end{algorithmic}
\end{algorithm}

\subsection{Security Analysis}

The privacy guarantee against the adversary threat model is captured by Proposition 1 below, while integrity (i.e., \ undetected bias of the aggregate) is captured by RB-LZKSA. Correctness is guaranteed from the antisymmetric mask cancellation~\eqref{eq:mask_cancel}, which yields $\sum_{r} y_r = \sum_{r} x_r$ exactly.

{\bf{Proposition 1}} (Information-Theoretic Privacy). Under the threat model, a coalition $\mathcal{C}$ of the coordinator and at most $R-2$ regions observes any honest region's input $x_r$ with per-observation mutual information bounded by $\tfrac{1}{2}\log_2(1 + \mathrm{Var}(x_r)/(|H|\sigma_m^2))$, where $|H| \geq 1$ is the number of remaining honest regions, provided the mask standard deviation is at least on the order of the data range, $\sigma_m \gtrsim \max_r|x_r|/10$.
%\end{proposition}

\begin{proof}
Let $r$ be an honest region and $\mathcal{C} \subseteq \{1,\ldots,R\} \setminus \{r\}$ with $|\mathcal{C}| \leq R-2$. By assumption at least one region $r_h \notin \mathcal{C} \cup \{r\}$ remains honest, so the mask component derived from $\kappa_{r,r_h}$ is unknown to $\mathcal{C}$.

Knowing $\{M_{r,r'}: r'\in \mathcal{C}\}$, the coalition reduces its observation to $\tilde{y}_r = y_r - \sum_{r' \in \mathcal{C}} M_{r,r'} = x_r + \tilde{m}_r$, where $\tilde{m}_r = \sum_{r' \in H} M_{r,r'} \sim \mathcal{N}(0, |H|\sigma_m^2)$ and $H = \{r' \notin \mathcal{C}\cup\{r\}\}$ has $|H|\geq 1$. The mutual information between $x_r$ and $\tilde{y}_r$ is therefore
\begin{equation}
I(x_r; \tilde{y}_r) \leq \tfrac{1}{2} \log_2\!\left(1 + \frac{\mathrm{Var}(x_r)}{|H|\,\sigma_m^2}\right).
\label{eq:mutual_info}
\end{equation}
Take $\sigma_m = 10^6$ and $|x_r| \leq 10^7$ for instance, the precondition $\sigma_m \gtrsim |x_r|/10$ holds, the SNR is $\leq 10^2$, and the per-observation leakage is bounded at $I \leq 3.3$~bits, which suffices to prevent single-shot recovery of the 64-bit value.

The RLWE commitments $\bm{c}_{r,j}$ are computationally hiding under decisional RLWE~\cite{lyubashevsky2010ideal}. Distinguishing them from uniform over $\Rq$ at $(n,q)=(1024,12289)$ requires more than 233 classical bits of work (NIST Level~3~\cite{nist2024fips203}) and is quantum-hard. Active adversarial deviations by $\mathcal{C}$ either fail Algorithm~\ref{alg:secure_agg}'s verification and abort the round (revealing nothing further about $x_r$) or leave the honest region's contribution intact (reducing to passive observation), so the same mutual-information bound applies in both cases.
\end{proof}

\subsection{Affine LP/QP Transformation}
\label{sec:affine}

Additive masking and reveal-bound commitment verification can collectively ensure the privacy and integrity of shared variables. However, accumulating cuts over hundreds of iterations could still leak structural information through the constraint matrix itself. To address this concern, the master is formulated in a secretly transformed variable space via an invertible affine map. The map is keyed by the pairwise Kyber-1024 shared secret established for masking in Section~\ref{sec:rlwe}, here in its role as the affine-map seed:
\begin{equation}
\kappa_{r,r'} = \mathsf{KA.Agree}(sk_r, pk_{r'}) = \mathsf{KA.Agree}(sk_{r'}, pk_r).
\label{eq:pairwise_key}
\end{equation}
A single group seed would require every region to hold the full map $(\bm{M},\bm{r})\in\R^{d\times d}\!\times\!\R^d$ (with $d = KT+1$) and thereby invert the transformation unilaterally. We instead build $\bm{M}$ and $\bm{r}$ row-by-row from pairwise secrets. For each tie-line $k$ connecting regions $r_1$ and $r_2$, the corresponding row is derived solely from $\kappa_{r_1,r_2}$:
\begin{align}
\bm{M}[\iota_{k,t},\, :] &= s_{k,t} \cdot \widehat{\text{PRF}}^{(M)}_{\kappa_{r_1,r_2}}(k,t), \label{eq:M_rows}\\
\bm{r}[\iota_{k,t}] &= \text{PRF}^{(r)}_{\kappa_{r_1,r_2}}(k,t). \nonumber
\end{align}
Here $\iota_{k,t}$ indexes the $(k,t)$ pair, the hat $\widehat{\,\cdot\,}$ denotes unit normalization of the PRF (pseudo-random function) output in $\R^d$, the superscripts $(M),(r),(\eta)$ are domain-separation labels (distinct roles invoke the same key $\kappa$ at distinct PRF instances), and $s_{k,t} \sim \text{Uniform}[0.5,\,2.0]$ (PRF-derived) controls the row norm. The epigraph row $\bm{M}_\eta$ is tie-line agnostic and constructed once via secure aggregation, $\bm{M}_\eta \propto \sum_{r<r'} \text{PRF}^{(\eta)}_{\kappa_{r,r'}}$. $\bm{M}$ is non-singular almost surely. Region $r$ derives only the rows corresponding to its own tie-lines $\mathcal{K}_r$ using its pairwise keys paired with adjacent neighbors and the shared $\bm{M}_\eta$:
\begin{equation}
\bm{M}_r \triangleq \bm{M}[\mathcal{K}_r \times \mathcal{T},\, :], \quad
\bm{M}_\eta \triangleq \bm{M}[d,\, :], \quad
\bm{r}_r \triangleq \bm{r}[\mathcal{K}_r \times \mathcal{T}].
\label{eq:partial_M}
\end{equation}
Rows for a non-adjacent tie-line $k'$ connecting $(r', r'')$ with $r \notin \{r', r''\}$ require $\kappa_{r',r''}$, which region $r$ does not hold. Non-adjacent data flows are therefore cryptographically hidden from region $r$, as region $r$ cannot recover $\bm{\theta}^*_{\mathcal{K}_{r'} \setminus \mathcal{K}_r}$ from $\bm{y}^*$, regardless of protocol compliance.

With the original variables $\bm{x} = [\bm{\theta}; \eta] \in \R^d$, the transformed substitution becomes
\begin{equation}
\bm{x} = \bm{M}\bm{y} + \bm{r}.
\label{eq:affine_map}
\end{equation}
Accordingly, the master LP relaxation without the proximal term takes the form:
\begin{align}
\min_{\bm{y}} \quad & \tilde{\bm{c}}^\top \bm{y} \label{eq:transformed_obj} \\
\text{s.t.} \quad & \tilde{\bm{A}} \bm{y} \leq \tilde{\bm{b}}, \label{eq:transformed_cons}
\end{align}
with transformed parameters
\begin{equation}
\tilde{\bm{c}} = \bm{M}^\top (\bm{c} + \bm{\delta}), \quad
\tilde{\bm{a}}_i = \bm{a}_i \bm{M}, \quad
\tilde{b}_i = b_i - \bm{a}_i \bm{r}.
\label{eq:transform_params}
\end{equation}
The small random perturbation $\bm{\delta} \in \R^d$, $\|\bm{\delta}\|_\infty \leq 10^{-6}$, prevents the coordinator from recovering $\bm{M}$ via the known sparsity pattern of $\bm{c}$. The transformed cost $\tilde{\bm{c}}$ in equation~\eqref{eq:transform_params} and perturbation $\bm{\delta}$ are assembled analogously. 

Rather than aggregating cut coefficients in $\bm{\theta}$-space and then transforming, each region transforms locally and the results are summed directly in $\bm{y}$-space. Since $\bm{\beta}_r[k,t] = 0$ for $k \notin \mathcal{K}_r$, the per-region transformed pair depends only on $\bm{M}_r$:
\begin{equation}
\tilde{\bm{a}}_r = \bm{\beta}_r^\top \bm{M}_r, \qquad
\tilde{b}_r = -(\alpha_r + \bm{\beta}_r^\top \bm{r}_r).
\label{eq:cut_transform}
\end{equation}
A designated region (e.g., region~1) absorbs the epigraph contribution via $\tilde{\bm{a}}_1 \mathrel{-}= \bm{M}_\eta$ and $\tilde{b}_1 \mathrel{+}= r_\eta$. The coordinator obtains the securely aggregated pair $\tilde{\bm{a}} = \sum_r \tilde{\bm{a}}_r$, $\tilde{b} = \sum_r \tilde{b}_r$ and appends $\tilde{\bm{a}} \bm{y} \leq \tilde{b}$ to the master. By linearity this matches the transform of the $\bm{\theta}$-space aggregate $(\alpha_{\text{agg}},\bm{\beta}_{\text{agg}})$, but no $\bm{\theta}$-space quantity ever reaches the coordinator.

\begin{figure}[t]
\centering
\begin{tikzpicture}[font=\tiny,
    nz/.style={draw, minimum width=8mm, minimum height=6mm, fill=blue!20, inner sep=0pt, font=\scriptsize},
    zr/.style={draw, minimum width=8mm, minimum height=6mm, fill=red!8, inner sep=0pt, font=\scriptsize, text=gray!50},
    slbl/.style={font=\tiny}
]
% ===== Left: Panel (a) z-space =====
\node[font=\scriptsize\bfseries] at (1.80, 3.75) {(a) $\bm{\theta}$-space (sparse)};

% Column headers (tie-lines)
\node[slbl] at (0.90, 3.30) {$k_{12}$};
\node[slbl] at (1.80, 3.30) {$k_{13}$};
\node[slbl] at (2.70, 3.30) {$k_{23}$};

% Row headers
\node[slbl, anchor=east] at (0.40, 2.80) {$R_1$};
\node[slbl, anchor=east] at (0.40, 2.10) {$R_2$};
\node[slbl, anchor=east] at (0.40, 1.40) {$R_3$};

% R1 row
\node[nz] at (0.90, 2.80) {2.5};
\node[nz] at (1.80, 2.80) {1.8};
\node[zr] at (2.70, 2.80) {0};
% R2 row
\node[nz] at (0.90, 2.10) {3.1};
\node[zr] at (1.80, 2.10) {0};
\node[nz] at (2.70, 2.10) {2.2};
% R3 row
\node[zr] at (0.90, 1.40) {0};
\node[nz] at (1.80, 1.40) {0.9};
\node[nz] at (2.70, 1.40) {1.7};

% Sum line
\draw[thick, red!70!black] (0.50, 0.95) -- (3.10, 0.95);
\node[slbl, anchor=east] at (0.40, 0.70) {$\bm{\beta}_{\text{agg}}$:};
\node[slbl, red!70!black] at (0.90, 0.70) {5.6};
\node[slbl, red!70!black] at (1.80, 0.70) {2.7};
\node[slbl, red!70!black] at (2.70, 0.70) {3.9};

% Vulnerability annotation
\draw[red!70!black, thick, decorate, decoration={brace, amplitude=2pt, mirror}]
    (0.50, 0.40) -- (3.10, 0.40)
    node[midway, below=3pt, font=\tiny, red!70!black]
    {2-party sum (vulnerable)};

% ===== Arrow (horizontal, rightward) =====
\draw[->, line width=0.8pt, blue!60!black] (3.25, 2.10) -- (4.00, 2.10);
\node[font=\tiny] at (3.625, 2.32) {$\bm{\beta}_r^{\top}\!\bm{M}_r$};
\node[font=\tiny\itshape, text=gray!45!black] at (3.625, 1.88) {dense PRF};

% ===== Right: Panel (b) y-space =====
\node[font=\scriptsize\bfseries] at (6.30, 3.75) {(b) $\bm{y}$-space (dense)};

% Column headers
\node[slbl] at (5.05, 3.30) {$y_1$};
\node[slbl] at (5.95, 3.30) {$y_2$};
\node[slbl] at (6.85, 3.30) {$y_3$};
\node[slbl] at (7.75, 3.30) {$y_4$};

% Row headers
\node[slbl, anchor=east] at (4.55, 2.80) {$R_1$};
\node[slbl, anchor=east] at (4.55, 2.10) {$R_2$};
\node[slbl, anchor=east] at (4.55, 1.40) {$R_3$};

% R1 row -- all non-zero
\node[nz] at (5.05, 2.80) {1.39};
\node[nz] at (5.95, 2.80) {0.69};
\node[nz] at (6.85, 2.80) {1.43};
\node[nz] at (7.75, 2.80) {0.10};
% R2 row -- all non-zero
\node[nz] at (5.05, 2.10) {3.05};
\node[nz] at (5.95, 2.10) {0.39};
\node[nz] at (6.85, 2.10) {0.01};
\node[nz] at (7.75, 2.10) {1.46};
% R3 row -- all non-zero
\node[nz] at (5.05, 1.40) {0.50};
\node[nz] at (5.95, 1.40) {1.74};
\node[nz] at (6.85, 1.40) {\scalebox{0.85}{$-1.10$}};
\node[nz] at (7.75, 1.40) {\scalebox{0.85}{$-0.28$}};

% Sum line
\draw[thick, blue!60!black] (4.65, 0.95) -- (8.15, 0.95);
\node[slbl, anchor=east] at (4.55, 0.70) {$\tilde{\bm{a}}$:};
\node[slbl, blue!60!black] at (5.05, 0.70) {4.94};
\node[slbl, blue!60!black] at (5.95, 0.70) {2.82};
\node[slbl, blue!60!black] at (6.85, 0.70) {0.34};
\node[slbl, blue!60!black] at (7.75, 0.70) {1.28};

% Protection annotation
\draw[blue!60!black, thick, decorate, decoration={brace, amplitude=2pt, mirror}]
    (4.65, 0.40) -- (8.15, 0.40)
    node[midway, below=3pt, font=\tiny, blue!60!black]
    {$R$-party sum (protected)};

\end{tikzpicture}
\caption{Transform-then-aggregate eliminates the per-tie-line two-party vulnerability. (a)~In $\bm{\theta}$-space, $\beta_r[k,t]=0$ for non-adjacent tie-lines, so each column of $\bm{\beta}_{\text{agg}}$ is only a two-party sum. (b)~Multiplying by dense PRF rows $\bm{M}_r$ converts every component of $\tilde{\bm{a}}_r$ to non-zero, so $\tilde{\bm{a}}{=}\sum_r \tilde{\bm{a}}_r$ becomes a genuine $R$-party sum protected by additive masking.}
\label{fig:transform_aggregate}
\vspace{-1em}
\end{figure}
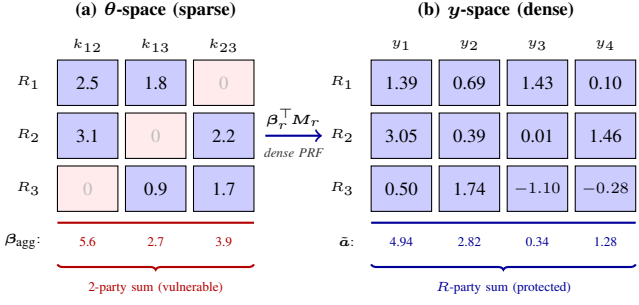

The ordering also avoids the per-tie-line sparsity vulnerability. Because $\bm{\beta}_r[k,t] = 0$ for $k \notin \mathcal{K}_r$, directly aggregating in $\bm{\theta}$-space would produce a two-party sum on any tie-line connecting $r_1$ and $r_2$, which is exposed as ADMM's pairwise consensus. In contrast, the multiplication $\tilde{\bm{a}}_r = \bm{\beta}_r^\top \bm{M}_r$ spreads each sparse $\bm{\beta}_r$ across all $d$ components of $\tilde{\bm{a}}_r$ via dense PRF rows, so every component of $\tilde{\bm{a}}$ is a genuine $R$-party sum. Fig.~\ref{fig:transform_aggregate} illustrates this with a numerical example. The same construction covers the LP-gradient evaluation that supplies $\bm{g}$ to the proximal-bundle quadratic form. Each region densifies its sparse gradient via
\begin{equation}
\tilde{\bm{g}}_r = \bm{M}^{\!\top}\!\!\begin{bmatrix}\nabla_{\bm{z}} C_r \\ 0\end{bmatrix} \in \R^{d},
\label{eq:grad_transform}
\end{equation}
and the coordinator aggregates $\tilde{\bm{g}} = \sum_r \tilde{\bm{g}}_r$, recovering $[\bm{g};\,0] = \bm{M}^{-\top}\tilde{\bm{g}}$ up to machine precision. Every component of $\tilde{\bm{g}}$ is a $R$-party sum, so the enhanced masking protection extends uniformly to cuts, costs, and gradients.

The coordinator sends $\bm{y}^*$ to each region $r$ via the encrypted communication channel, and region $r$ recovers only its own tie-line flows:
\begin{equation}
\bm{\theta}^*_{\mathcal{K}_r} = \bm{M}_r \bm{y}^* + \bm{r}_r.
\label{eq:region_recovery}
\end{equation}
Absent $\bm{M}_{r'}$ for $r' \neq r$, region $r$ cannot compute $\bm{\theta}^*_{\mathcal{K}_{r'}}$, achieving cryptographic data flow isolation between non-adjacent regions. The proximal term in the master QP expands as
$(\mu/2)\norm{\bm{\theta}-\hat{\bm{\theta}}}^2 = (\mu/2)\bm{y}^\top \bm{Q}\bm{y} + \bm{g}^\top\bm{y} + \mathrm{const.}$,
where the Gram matrix $\bm{Q} = \bm{M}_{1:KT}^\top \bm{M}_{1:KT}$ is built once at setup by securely aggregating the per-region outer products $\bm{M}_r^\top \bm{M}_r$, and the linear term $\bm{g}$ is refreshed each iteration by securely aggregating $\bm{M}_r^\top(\bm{M}_r \hat{\bm{y}} + \bm{r}_r - \hat{\bm{\theta}}_{\mathcal{K}_r})$. The coordinator therefore solves the proximal quadratic programming without holding $\bm{M}$ or $\hat{\bm{\theta}}$.

All inter-party messages, i.e., masked values, RLWE commitments, master data, and per-iteration $\bm{y}^*$ vectors, are transmitted over region-specific channels secured by post-quantum authenticated encryption (per-pair session keys via the same primitive used for $\kappa_{r,r'}$ in equation~\eqref{eq:pairwise_key}), providing information security under the Module-LWE hardness~\cite{nist2024fips203, bos2018crystals}. Fig.~\ref{fig:privacy_arch} illustrates the overall architecture, where each region submits encrypted values through the PQ-secure aggregation layer and the coordinator receives only the aggregated sum.

% ============================================================================
\section{Simulation Results}
\label{sec:results}
% ============================================================================

\subsection{Simulation Setup}

We evaluate the proposed approach on three benchmarks, including the IEEE 118-bus, 300-bus, and ACTIVSg500 systems, from MATPOWER 8.0~\cite{zimmerman2010matpower}. All experiments are conducted on an Intel Core Ultra 7 165H with 16 cores, allied with the Gurobi 12.0.2 and 8 solver threads. The Benders approach uses a 0.1\% gap tolerance, 250 maximum iterations, and time limits of 600\,s (IEEE-118/300) and 900\,s (ACTIVSg500). Proximal bundle parameters: $\mu_0 = 5.0$, serious step threshold $\rho_s = 0.05$, maximum 800 active cuts with 30-iteration age limit. All Benders results are averaged over five independent trials with different random seeds to quantify MIP solver non-determinism arising from Gurobi's multi-threaded branch-and-bound. Four methods are compared to isolate the effects of decomposition structure and privacy, i.e., \emph{Centralized MIP}: monolithic Gurobi solve providing the optimal reference cost $c^\star$; \emph{ADMM Consensus}: a widely used distributed baseline~\cite{boyd2011distributed, erseghe2014distributed} providing primal solutions without certified bounds or PQ privacy; \emph{Benders (privacy disabled)}: the proposed distributed algorithm with all privacy layers deactivated; \emph{PQ-Benders}: the proposed distributed optimization with full privacy preservation (additive sharing, RLWE commitments, affine transformation, PQ-encrypted channels).

\subsection{Solution Optimality}

\begin{figure}[!tb]
	\centering
	\includegraphics[width=0.5\textwidth]{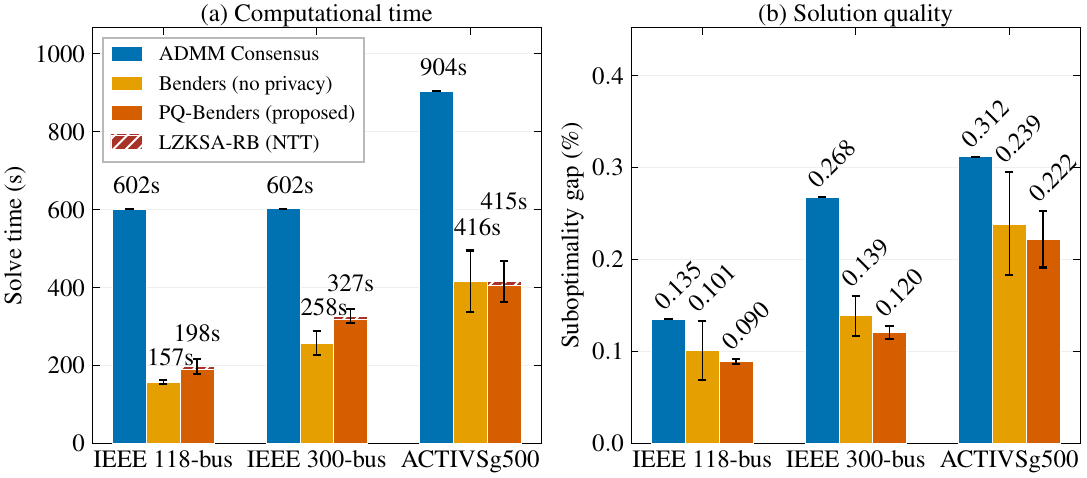}
	\caption{Method comparison across IEEE-118, IEEE-300, and ACTIVSg500. (a)~Computational time. (b)~Solution quality.}
	\label{fig:method_comparison}
	\vspace{-1em}
\end{figure}

All gaps are measured uniformly as $(UB - c^\star) / c^\star \times 100$, where $c^\star$ is the optimal cost from centralized approach. ADMM is given the same time budget as Benders (600-900\,s) with zero convergence tolerance. As in Fig. \ref{fig:method_comparison}, PQ-Benders achieves mean suboptimality of $0.090\pm0.003\%$, $0.120\pm0.007\%$, and $0.222\pm0.031\%$ on three systems, respectively, while ADMM Consensus corresponding derive gaps of $0.158\%$, $0.307\%$, and $0.313\%$. PQ-Benders dominates ADMM on all three systems under the same metric and additionally provides post-quantum privacy preservation that ADMM lacks. Multi-trial experiments further reveal that ADMM is deterministic across five independent random seeds. This structural floor stems from ADMM's pairwise consensus update and its convergence is guaranteed on convexity of the regional subproblems. When subproblems contain integer variables, the convex analysis prerequisite is violated, the algorithm may oscillate or stall at a local fixed point, and the optimality gap admits no algorithmic mechanism for closure. Compared with mean gaps of Benders (privacy disabled), i.e., $0.139\%$, $0.222\%$ and $0.239\%$, PQ-Benders' mean gaps are slightly lower. In all cases, confidence intervals overlap substantially. This is mainly attributable to MIP solver non-determinism. Gurobi's multi-threaded branch-and-bound possibly explores different subtrees depending on thread scheduling, and the affine transformation alters the LP's numerical pivoting sequence, leading to different convergence paths across iterations.

\subsection{Convergence and Cryptographic Overhead}
\begin{figure}[!tb]
\centering
\includegraphics[width=0.5\textwidth]{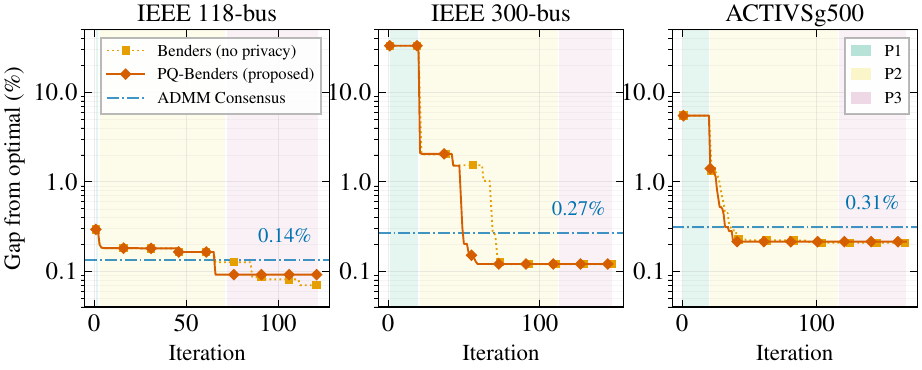}
\caption{Upper-bound convergence trajectories for PQ-Benders (solid) and Benders without privacy (dotted) on IEEE-118, IEEE-300 , and ACTIVSg500. The ADMM Consensus final suboptimality (dash-dot) is shown as a horizontal reference. }
\label{fig:convergence}
\end{figure}

\begin{figure}[!tb]
\centering
\includegraphics[width=\columnwidth]{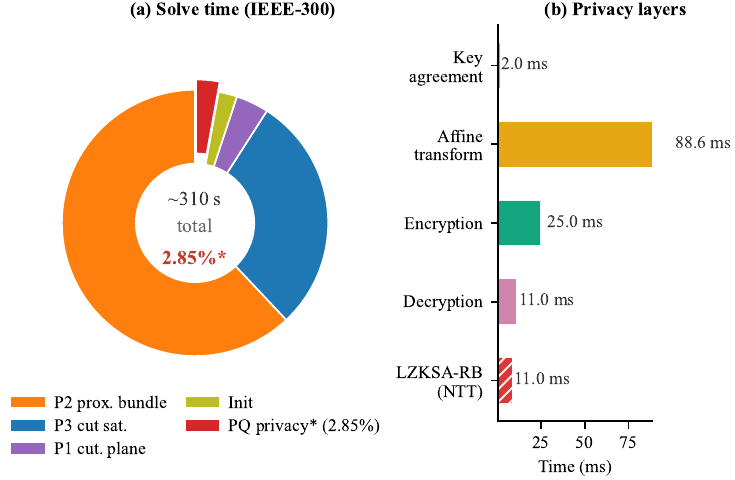}
\caption{Privacy overhead analysis (IEEE-300, trial 0), with both panels reporting RB-LZKSA (with NTT) cost: $R{=}3$, $\lambda{=}8$, sequential verifier; $n{=}1024$, $q{=}12289$. (a)~Breakdown of total solve time. (b)~Privacy layer component costs.}
\label{fig:privacy_overhead}
\vspace{-1em}
\end{figure}

Fig.~\ref{fig:convergence} gives the convergence trajectory of the three-phase convergence strategy (P1: initial LP relaxation, P2: commitment refinement, P3: proximal tightening), which is employed in Algorithm~\ref{alg:benders}. Phase~1 rapidly tightens the initial gap in 2-20 iterations via multi-point cutting planes with a coarse MIP (1\% gap). Phase~2 activates proximal stabilization over 65-104 iterations, confining the master solution to a trust region via $(\mu/2)\norm{\bm{\theta} - \hat{\bm{\theta}}}^2$ with a tighter oracle (0.2\% gap), eliminating the extreme-point oscillation that otherwise generates locally sharp but globally redundant cuts. Phase~3 targets the weakest polyhedral approximation region by generating cuts at $\bm{\theta}^{\text{lb}}$ rather than $\bm{\theta}^*$ with the tightest oracle (0.05\% gap), sustaining lower-bound improvement over 111-146 iterations. The adaptive $\mu$ schedule automates the transition between exploration and refinement. PQ-Benders and the privacy-disabled version exhibit nearly identical convergence trajectories across all 5 trials per method, confirming that the privacy preservation mechanism, including the affine transformation, preserves numerical stability and introduces no systematic performance degradation. Fig.~\ref{fig:privacy_overhead} visualizes the deployed cryptographic components alongside the RB-LZKSA (with Number Theoretic Transform (NTT)) layer ($\sim$$8.7$\,s on IEEE-300, which is empirically measured $11.89$ times speedup at $n=1024$ compared to the classic version). The overall privacy preservation cost takes up 2.1-3.8\% of solve time across the three systems.

\subsection{Comparison with Privacy-Preserving Noisy ADMM}
\label{sec:noisy_admm_comparison}

Optimality comparisons between the classic ADMM Consensus and PQ-Benders are conducted. To further assess the privacy preservation performance, ADMM combined with the privacy-preserving noise mechanism is considered. To be aligned with the canonical privacy-aware distributed optimization approaches~\cite{han2016differentially, dvorkin2020differentially}, we therefore implement a multiplicative-noise variant of ADMM, i.e., noisy ADMM, in which each pairwise consensus update is perturbed by the Homomorphic encryption-compatible multiplicative Gaussian noise \cite{liu2023privacy}:
\begin{equation}
\bar{\theta}_{ab}^{(k)} \;=\; \tfrac{1}{2}\,\xi^{(k)}\!\left(\theta_a^{(k)} + \theta_b^{(k)}\right),
\qquad
\xi^{(k)} \sim \mathcal{N}\!\left(1,\,\sigma_k^{\,2}\right),
\label{eq:noisy_admm}
\end{equation}
with a phase-aware decay schedule $\sigma_k = \sigma_0 / (1+k)^{0.5}$ during the LP-warmup phase, reduced by a MIP-noise fraction of 0.3 to $\sigma_k = 0.3 \times \sigma_0 / (1+k)^{0.5}$ for $k \ge 30$ once the regional subproblems become MIP. We sweep three noise intensities, $\sigma_0 \in \{0.05, 0.10, 0.20\}$, corresponding respectively to mild, moderate, and strong privacy settings. For each pair $(\sigma_0)$, we run five independent trials under the same time budget as PQ-Benders, tracking both a vanilla upper bound (lowest upper bound achieved at the noisy $\bar{\theta}$).

\begin{figure}[!tb]
\centering
\includegraphics[width=0.5\textwidth]{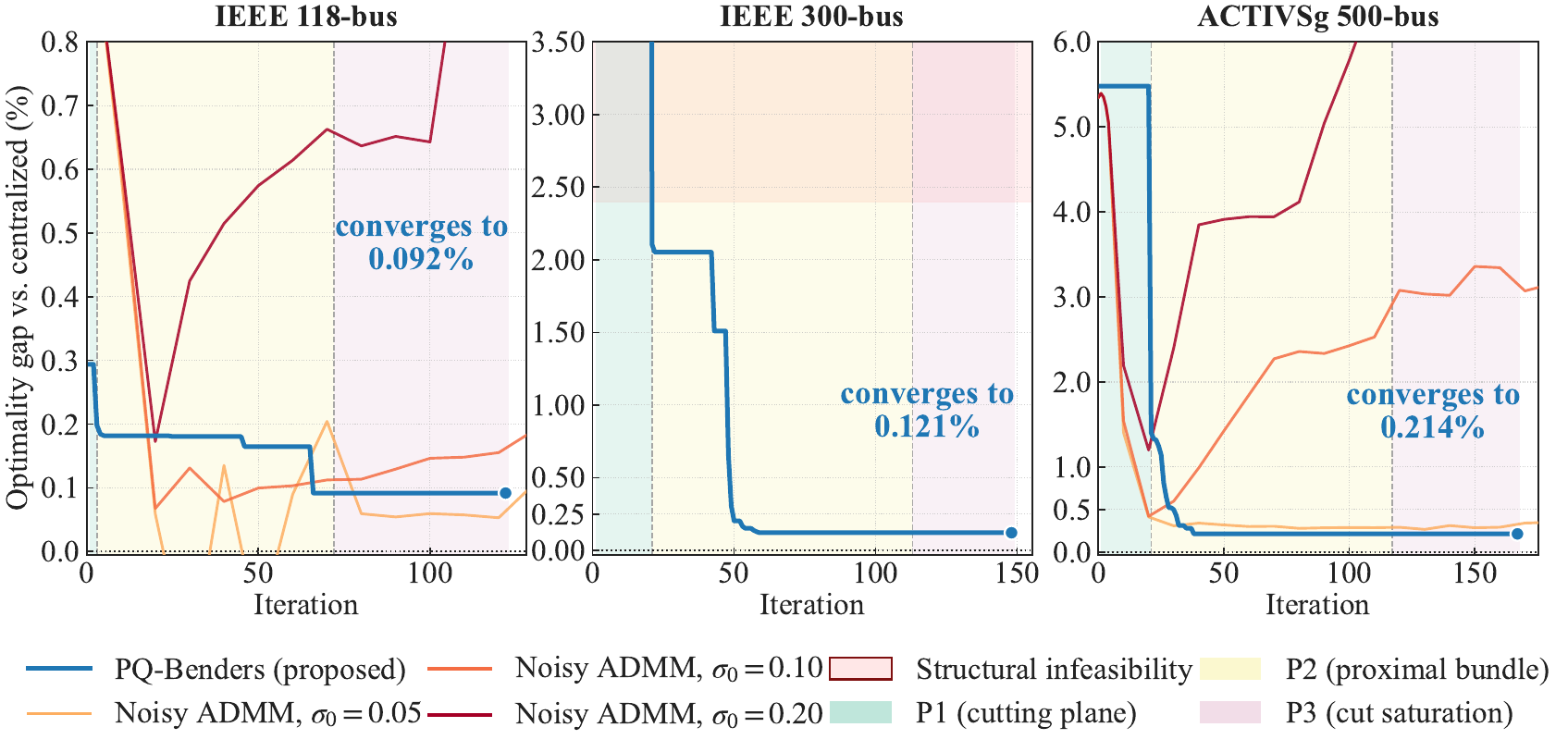}
\caption{Iteration-wise centralized-gap convergence of PQ-Benders (thick blue) and privacy-preserving ADMM (three noise levels, $\sigma_0 \in \{0.05, 0.10, 0.20\}$) on the three systems.}
\label{fig:noisy_admm_convergence}
%\vspace{-1.5em}
\end{figure}

Fig.~\ref{fig:noisy_admm_convergence} traces the iteration-wise trajectories of PQ-Benders and noisy ADMM, shown by the trial-0 trajectories. On three systems, PQ-Benders descends monotonically through its three phases, converging within 120-170 iterations to centralized gaps. The noisy ADMM trajectories, in contrast, either stagnate at a noise-dependent floor (ACTIVSg500, $\sigma_0\!=\!0.20$, and $1.5$-$4.4\%$ across different trials) or exhibit persistent oscillations around a biased mean (IEEE-118, $\sigma_0\!=\!0.20$, and the curve fails to settle below $0.5\%$). On IEEE-300 the situation is qualitatively worse: every noisy ADMM trial yields a commitment schedule that is structurally infeasible, producing a rescued-UB gap of $\approx$$78{,}000\%$. This 100\% infeasibility rate is a direct consequence of the discrete commitment grid, which contains large dead zones in which small multiplicative perturbations of $\bar{\theta}$ fall outside the feasible lattice of $\{0,1\}$ assignments. PQ-Benders avoids this failure mode entirely because its additive secret-sharing layer produces a zero-error sum~\eqref{eq:mask_cancel}, so the master MIP always sees a feasible $\bm{\theta}$.

\begin{figure}[!tb]
\centering
\includegraphics[width=0.5\textwidth]{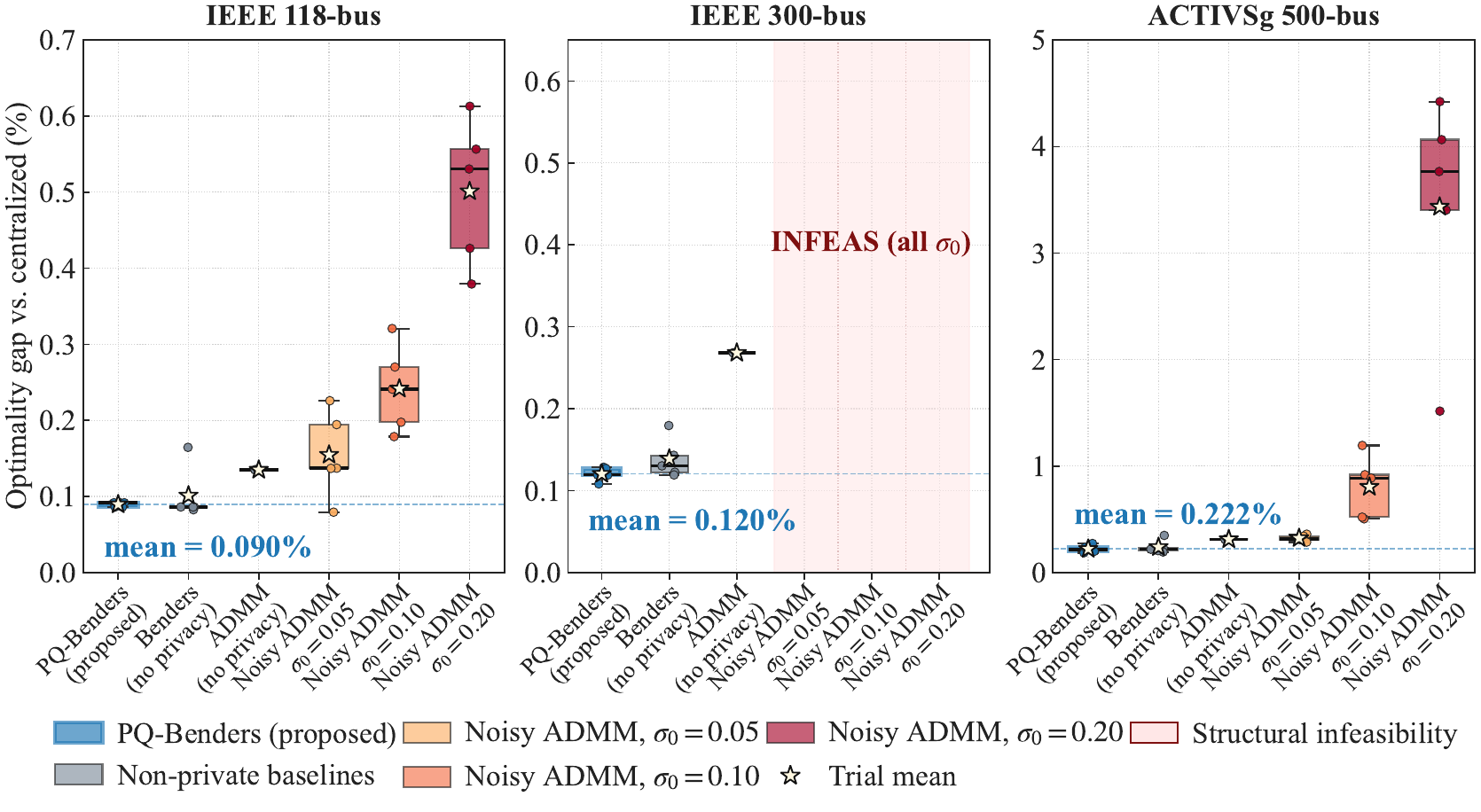}
\caption{Optimality gap distribution (5 trials per method per system). Stars ($\star$) mark trial means.}
\label{fig:noisy_admm_boxplots}
%\vspace{-1.5em}
\end{figure}

Meanwhile, PQ-Benders' trial-to-trial variability is very small (standard deviations of $0.003$/$0.007$/$0.058$\% on three systems), comparable to the privacy-disabled Benders ($0.032$/$0.022$/$0.056$\%), as in Fig.~\ref{fig:noisy_admm_boxplots}. In contrast, noisy ADMM exhibits a monotone gap-noise trade-off. On ACTIVSg500 the mean gap inflates from $0.322\%$ at $\sigma_0\!=\!0.05$ to $0.805\%$ at $\sigma_0\!=\!0.10$ and then to $3.435\%$ at $\sigma_0\!=\!0.20$, with inter-quartile ranges that are one to two orders of magnitude wider than PQ-Benders' and outlying trials reaching $4.4\%$. Even at the mildest noise setting ($\sigma_0\!=\!0.05$), noisy ADMM fails to dominate PQ-Benders on IEEE-118. Its $0.155\%$ mean is much worse than the cutting-plane PQ-Benders mean of $0.090\%$. On ACTIVSg500 the  mean at $\sigma_0\!=\!0.05$ ($0.322\%$) is $1.45$ times worse than the cutting-plane PQ-Benders mean ($0.222\%$), and at higher noise levels noisy ADMM degrades by up to $15$ times. On IEEE-300, noisy ADMM remains catastrophically infeasible at every noise level tested, exposing the incompatibility between multiplicative noise and the binary commitment lattice and ruling it out as a practical privacy mechanism for the MIP-based unit commitment. PQ-Benders, by contrast, is near optimal by construction and remains stable across all three test systems.

\subsection{Reserve Sharing Sensitivity}

\begin{figure*}[!tb]
\centering
\includegraphics[width=0.8\textwidth]{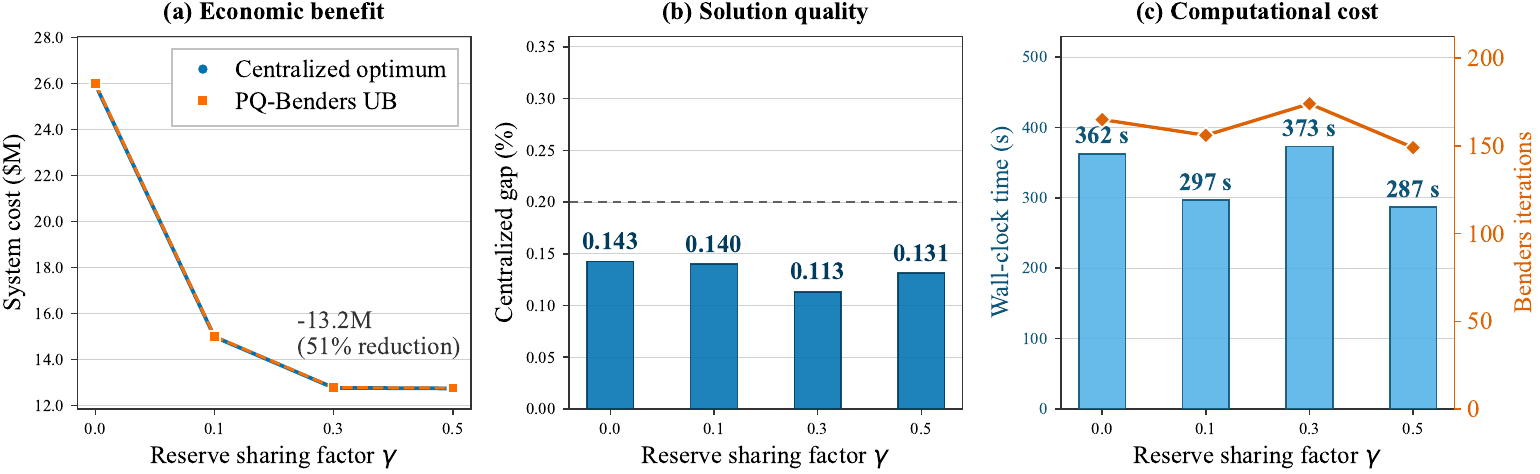}
\caption{Reserve sharing sensitivity of the proposed PQ-Benders on IEEE-300 with $\gamma\in\{0,0.1,0.3,0.5\}$. (a)~Economic benefit. (b)~Solution quality. (c)~Computational cost.}
\label{fig:reserve_sens}
\vspace{-1.2em}
\end{figure*}

The inter-regional reserve sharing factor $\gamma$ (equations~\eqref{eq:res_up}-\eqref{eq:res_dn}) is a policy parameter that grid operators can adjust to balance reliability against economic efficiency. Fig.~\ref{fig:reserve_sens} examines its impact on the IEEE-300 system using the proposed PQ-Benders solver approach over a four-point sweep $\gamma \in \{0, 0.1, 0.3, 0.5\}$ at a uniform $600\,\text{s}$ time budget per run.

As in Fig.~\ref{fig:reserve_sens}(a), enabling inter-regional reserve-sharing through unused tie-line capacity yields an obvious cost reduction. The centralized optimum drops from $\$25.97$M at $\gamma=0$ (isolated reserves) to $\$12.73$M at $\gamma=0.5$, a $\$13.24$M ($51\%$) reduction. Most of the gain is realized by modest sharing ($\gamma=0.1$ already captures over $80\%$ of the total savings), with diminishing returns beyond $\gamma=0.3$. Meanwhile, the PQ-Benders gap $(UB-c^\star)/c^\star$ stays comfortably below the $0.2\%$ for every tested $\gamma$, ranging from $0.113\%$ at $\gamma=0.3$ to at most $0.143\%$ at $\gamma=0$, as in Fig.~\ref{fig:reserve_sens}(b). Within the operationally relevant region $\gamma\in[0.3,0.5]$ the gap remains tight at $0.113\%$ and $0.131\%$ for $\gamma=0.3$ and $\gamma=0.5$ respectively, which indicates the proposed approach performs stable under different reserve-sharing policies.

As in Fig.~\ref{fig:reserve_sens}(c), the proposed PQ-Benders approach requires $287$-$373\,\text{s}$ ($148$-$173$ iterations) to derive near-optimal solutions across all factors ($\gamma$), with no monotone dependence on the sharing intensity. Per-iteration efficiency remains around ($1.9$-$2.2\,\text{s/iter}$). Overall, these results verify that the proposed PQ-Benders approach is simultaneously effective (capturing the full $51\%$ economic benefit of reserve sharing), accurate (optimality gap $\le 0.143\%$), and robust (stable per-iteration efficiency) across the reserve-sharing policies.

% ============================================================================
\section{Conclusion}
\label{sec:conclusion}
% ============================================================================
Multi-region unit commitment with reserve sharing is computationally and institutionally challenging because it involves mixed-integer regional subproblems and the exchange of sensitive operational data across regions. As post-quantum security becomes increasingly important for critical infrastructure, quantum-resilient privacy protection is expected to become an operational requirement rather than a desirable feature. To address these challenges, this paper proposes a post-quantum-secure Benders-based distributed optimization approach. By leveraging the global summation structure of customized Benders decomposition, the proposed approach admits additive secret sharing, affine variable transformation, and reveal-bound commitment verification to deliver post-quantum information-theoretic privacy at no more than $3.8\%$ computational overhead of total solve time, whereas ADMM's pairwise coupling is inherently incompatible with multi-party privacy guarantees. Simulation results on three systems verify that the proposed approach achieves mean suboptimality of $0.09\%$-$0.22\%$, which are $1.4$-$2.6$ times tighter than ADMM Consensus. The proposed approach imposes no measurable cost-quality trade-off relative to its privacy-disabled counterpart, whereas noisy ADMM degrades monotonically under tightening privacy budgets and is structurally infeasible on the IEEE 300-bus system. Inter-regional reserve sharing recovers up to $51\%$ of system cost, realized by the proposed approach across the reserve-sharing policies. Future work will extend the proposed approach to stochastic multi-period dispatch under renewable uncertainty.

\bibliographystyle{IEEEtran}
\bibliography{paper_TSG}
\end{document}